\documentclass[a4paper,12pt]{article}
\usepackage[utf8x]{inputenc}
\usepackage{amsfonts}
\usepackage{latexsym}
\usepackage{amsmath}
\usepackage{amssymb}
\usepackage{amssymb}

\usepackage{slashed}
\usepackage{upgreek}
\usepackage{mathrsfs}
\usepackage{bbm}

\usepackage{relsize}
\usepackage{graphicx}

\numberwithin{equation}{section}

\newcommand{\newsection}{    
						\setcounter{equation}{0}
						\section
					}
\def\appendix#1{	\addtocounter{section}{1}
				\setcounter{equation}{0}
				\renewcommand{\thesection}{\Alph{section}}
				\section*{Appendix \thesection\protect\indent \parbox[t]{11.15cm}{#1}}
				\addcontentsline{toc}{section}{Appendix \thesection\ \ \ #1}
			}

\newcommand{\be}{\begin{eqnarray}}
\newcommand{\ee}{\end{eqnarray}}
\newcommand{\bea}{\begin{eqnarray}}
\newcommand{\eea}{\end{eqnarray}}
\newcommand{\ba}{\begin{array}}
\newcommand{\ea}{\end{array}}

\newenvironment{frcseries}{\fontfamily{frc}\selectfont}{}

\def \la {\label}

\def\e{\epsilon}

\def\bbe{{\bf{e}}}
\font\mybb=msbm10 at 11pt

\def\bb#1{\hbox{\mybb#1}}

\def\bR {\bb{R}}

\def\tn {{\tilde{\nabla}}}

\begin{document}
\begin{titlepage}
\begin{center}
\vspace*{-1.0cm}
\hfill DMUS--MP--14/15 \\

\vspace{2.0cm} {\Large \bf Dynamical symmetry enhancement near massive IIA horizons} \\[.2cm]

\vspace{1.5cm}
 {\large  U.~Gran$^1$, J.~Gutowski$^2$, U.~Kayani$^3$ and  G.Papadopoulos$^3$}

\vspace{0.5cm}

${}^1$ Fundamental Physics\\
Chalmers University of Technology\\
SE-412 96 G\"oteborg, Sweden\\

\vspace{0.5cm}
$^2$ Department of Mathematics \\
University of Surrey \\
Guildford, GU2 7XH, UK \\

\vspace{0.5cm}
${}^3$ Department of Mathematics\\
King's College London\\
Strand\\
London WC2R 2LS, UK\\

\vspace{0.5cm}

\end{center}

\vskip 1.5 cm
\begin{abstract}

\vskip1cm

We prove that Killing horizons in massive IIA supergravity
 preserve an even number of supersymmetries,
and that their symmetry algebra  contains an  $\mathfrak{sl}(2, \bR)$ subalgebra, confirming the conjecture of \cite{iibindex}.
We also prove a new class of  Lichnerowicz type theorems
for connections of the spin bundle whose holonomy is contained in a general linear group.

\end{abstract}

\end{titlepage}



\setcounter{section}{0}
\setcounter{subsection}{0}
\setcounter{equation}{0}

\newsection{Introduction}

It has been known for some time that there is (super)symmetry enhancement near black hole and brane horizons. This has been
observed on a case by case basis, see e.g.~\cite{carter, gibbons1, gwgpkt}, and it has been extensively used in the
development of the AdS/CFT correspondence \cite{maldacena}. Recently, it has been realized  that the (super)symmetry enhancement near Killing horizons is a generic phenomenon
which depends only on the smoothness of the fields and some global assumptions on the spatial horizon sections.
The concise conjecture has been stated in \cite{iibindex}, following some earlier results
in \cite{5index} and \cite{11index}. This conjecture includes all the (super)symmetry enhancement phenomena near black hole Killing horizons
as special cases.  So far the conjecture has been verified in a variety of theories which
include the minimal 5-dimensional gauged supergravity, M-theory, and IIB and IIA supergravities \cite{5index, iibindex, 11index, iiaindex}.

In this paper, we shall prove the conjecture of \cite{iibindex} for the massive IIA horizons, i.e.~the Killing horizons of massive IIA supergravity. This in particular implies that
massive IIA horizons with smooth fields and spatial horizon sections, ${\cal S}$, which are compact without boundary:

\begin{itemize}

\item Preserve an even number of supersymmetries
\bea
N=2 N_- ~,
\eea
where $N_-$ is the dimension of the kernel of a Dirac like operator ${\mathscr D}^{(-)}$ on ${\cal S}$ which depends on the fluxes.

\item The symmetry group of all such horizons
contains an $\mathfrak{sl}(2, \bR)$ subalgebra.

\end{itemize}

The proof of the conjecture for massive IIA horizons is similar to that given in \cite{iiaindex} for standard IIA horizons but there is a key difference.
Massive IIA supergravity has a negative cosmological constant. The proof of the conjecture relies on the application of the maximum principle
to demonstrate certain Lichnerowicz type theorems. In turn the application of the maximum principle requires the positive semi-definiteness of a certain
term which depends the fluxes. The existence of a negative cosmological constant in the theory has the potential of invalidating the
arguments based on the maximum principle as it can contribute with the opposite sign in the expressions required for the application of the maximum principle.
We show that this is not the case and therefore the conjecture can be extended to massive IIA horizons.

Nevertheless many of the steps in the proof of the conjecture for massive IIA horizons are similar to those presented for IIA horizons
in \cite{iiaindex}.  Because of this, in the main body of the paper, we shall state the key statements and formulae required for the proof of the conjecture.  The detailed proofs
of these are presented in the appendices.

This paper is organised as follows. In section 2, we show that massive IIA horizons preserve an even number of supersymmetries. In section 3, we demonstrate
that the symmetry of massive IIA horizons includes an $\mathfrak{sl}(2, \bR)$ subalgebra. In addition, in appendix A, we give the field equations
of the near horizon fields. In appendix B, we identify the independent KSEs of the near horizon geometries. In appendix C, we derive some key formulae
which are required for the proof of Lichnerowicz type theorems for ${\mathscr D}^{(\pm)}$ operators, and in appendix D we present some identities which are necessary to
demonstrate the $\mathfrak{sl}(2, \bR)$ invariance of massive IIA horizons.

\newsection{Supersymmetry enhancement}

\subsection{Independent KSEs}

The first part of the conjecture states that massive IIA horizons preserve an even number of supersymmetries. In particular, if the massive IIA horizons admit one supersymmetry,
then this enhances to two. To prove this, we solve the KSEs of massive IIA supergravity \cite{romans}
\bea
{\cal D}_\mu\e&\equiv&\nabla_\mu\e +{1\over8} H_{\mu\nu_1\nu_2} \Gamma^{\nu_1\nu_2}\Gamma_{11}\e+{1\over16} e^\Phi F_{\nu_1\nu_2} \Gamma^{\nu_1\nu_2} \Gamma_\mu \Gamma_{11} \e
\cr
&& +{1\over 8\cdot 4!}e^\Phi G_{\nu_1\nu_2\nu_3\nu_4} \Gamma^{\nu_1\nu_2\nu_3\nu_4} \Gamma_\mu\e +\frac{1}{8}e^{\Phi}m\Gamma_{\mu}\e=0~,\label{GKSE}\\
{\cal A}\e&\equiv&\partial_\mu\Phi\, \Gamma^\mu \e+{1\over12} H_{\mu_1\mu_2\mu_3} \Gamma^{\mu_1\mu_2\mu_3} \Gamma_{11} \e+{3\over8} e^\Phi F_{\mu_1\mu_2} \Gamma^{\mu_1\mu_2} \Gamma_{11} \e
\cr
&&+ {1\over 4\cdot 4!}e^\Phi\, G_{\mu_1\mu_2\mu_3\mu_4} \Gamma^{\mu_1\mu_2\mu_3\mu_4}\e + \frac{5}{4}e^{\Phi}m \e=0~,\label{AKSE}
\eea
for the near horizon fields
\be
ds^2 &=&2 \bbe^+ \bbe^- + \delta_{ij} \bbe^i \bbe^j~,~~~G= \bbe^+ \wedge \bbe^- \wedge X +r \bbe^+ \wedge Y + \tilde G~,~~
\cr
H &=&\bbe^+ \wedge \bbe^- \wedge L+ r \bbe^+ \wedge M + \tilde H~,~~~F= \bbe^+ \wedge \bbe^- S + r \bbe^+ \wedge T+ \tilde F~,
\la{hormetr}
\ee
where $\epsilon$ is a commuting Majorana $Spin(9,1)$ spinor and we have introduced the frame
\be
\label{basis1}
\bbe^+ = du, \qquad \bbe^- = dr + rh -{1 \over 2} r^2 \Delta du, \qquad \bbe^i = e^i_I dy^I~.
\ee
This expression for the near horizon fields is similar to that for the IIA case in \cite{iiaindex} though their dependence on the gauge potentials is different.  The massive theory contains an additional parameter $m$, the mass term, and the fields and both the gravitino and dilatino KSEs depend on it, see appendix A. Furthermore, the Bianchi identities relate some of the components of the near horizon fields. In particular, $M$, $T$ and $Y$ are not independent, see again appendix A. The dependence on the coordinates $u,r$ is given explicitly and all the fields depend on the coordinates $y^I$
of the spatial horizon section ${\cal S}$ defined by $u=r=0$.

The KSEs of massive IIA supergravity can be solved along the lightcone directions. The solution is
\bea\label{lightconesol}
\e=\e_++\e_-~,~~~\e_+=\phi_+(u,y)~,~~~\e_-=\phi_-+r \Gamma_-\Theta_+ \phi_+~,
\eea
and
\bea
\phi_-=\eta_-~,~~~\phi_+=\eta_++ u \Gamma_+ \Theta_-\eta_-~,
\eea
where
\bea
\Theta_\pm &=& {1\over4} h_i\Gamma^i\mp{1\over4} \Gamma_{11} L_i \Gamma^i-{1\over16} e^{\Phi} \Gamma_{11} (\pm 2 S+\tilde F_{ij} \Gamma^{ij})
\cr
&-& {1\over8 \cdot 4!} e^{\Phi} (\pm 12 X_{ij} \Gamma^{ij}
+\tilde G_{ijkl} \Gamma^{ijkl}) - \frac{1}{8}e^{\Phi}m,
\nonumber \\
\eea
 $\Gamma_\pm\epsilon_\pm=0$, and $\eta_\pm=\eta_\pm(y)$ depend only on the coordinates $y$ of the spatial horizon section ${\cal S}$. Both $\eta_\pm$ are sections of the $Spin(8)$ bundle over ${\cal S}$ associated
  with the Majorana representation.

Substituting the spinor $\epsilon$ given in (\ref{lightconesol}) into the KSEs (\ref{GKSE}) and (\ref{AKSE}),  one obtains a large number of conditions given in appendix B. To describe the remaining independent KSEs consider the operators
\bea
\nabla_{i}^{(\pm)}&=& \tilde{\nabla}_{i} + \Psi^{(\pm)}_{i}~,
\eea
with
\bea
\label{alg1pm}
\Psi^{(\pm)}_{i} &=& \bigg( \mp \frac{1}{4}h_{i} \mp \frac{1}{16}e^{\Phi}X_{l_1 l_2}\Gamma^{l_1 l_2}\Gamma_{i} + \frac{1}{8 \cdot 4!}e^{\Phi}{\tilde{G}}_{l_1 l_2 l_3 l_4}\Gamma^{l_1 l_2 l_3 l_4}\Gamma_{i} + \frac{1}{8}e^{\Phi}m \Gamma_i \bigg)
\cr
&+& \Gamma_{11}\bigg(\mp \frac{1}{4}L_{i} + \frac{1}{8}{\tilde{H}}_{i l_1 l_2}\Gamma^{l_1 l_2}
\pm \frac{1}{8}e^{\Phi}S\Gamma_{i} - \frac{1}{16}e^{\Phi}{\tilde{F}}_{l_1 l_2}\Gamma^{l_1 l_2}\Gamma_{i}\bigg)~,
\eea
and
\bea
\label{alg2pm}
\mathcal{A}^{(\pm)} &=& \partial_i \Phi \Gamma^i  + \bigg(\mp \frac{1}{8}e^{\Phi}X_{l_1 l_2}\Gamma^{l_1 l_2} + \frac{1}{4 \cdot 4!}e^{\Phi}{\tilde{G}}_{l_1 l_2 l_3 l_4}\Gamma^{l_1 l_2 l_3 l_4} + \frac{5}{4}e^{\Phi}m \bigg)
\cr
&+& \Gamma_{11}\bigg(\pm \frac{1}{2}L_{i}\Gamma^{i} - \frac{1}{12}{\tilde{H}}_{i j k}\Gamma^{i j k} \mp \frac{3}{4}e^{\Phi}S + \frac{3}{8}e^{\Phi}{\tilde{F}}_{i j}\Gamma^{i j}\bigg)~.
\eea
These are derived from the naive restriction of the supercovariant derivative and the dilatino KSE on ${\cal S}$.
\vskip 0.3cm
{\it Theorem:} The remaining independent KSEs are
\bea
\label{covr}
\nabla_{i}^{(\pm)}\eta_{\pm}  = 0~,~~~\mathcal{A}^{(\pm)}\eta_{\pm} = 0~.
\eea
Moreover if $\eta_-$ solves the KSEs, then
\bea
\eta_+ = \Gamma_{+}\Theta_{-}\eta_{-}~,
\label{epfem}
\eea
is also a solution.

\vskip 0.3cm

{\it Proof:} The proof is given in appendix B.

\rightline{ $\square$}

\subsection{Lichnerowicz type theorems for ${\mathscr D}^{(\pm)}$ }
\label{lichthx}

To proceed with the proof of the first part of the conjecture define the modified horizon Dirac operators   as
\bea
\label{redef2}
{\mathscr D}^{(\pm)}={\cal D}^{(\pm)} -{\cal A}^{(\pm)}~,
\eea
 where
\bea
{\cal D}^{(\pm)} \equiv \Gamma^{i}\nabla_{i}^{(\pm)} = \Gamma^{i}\tilde{\nabla}_{i} + \Psi^{(\pm)}~,
\eea
with
\bea
\label{alg3pm}
\Psi^{(\pm)} \equiv \Gamma^{i}\Psi^{(\pm)}_{i} &=& \mp\frac{1}{4}h_{i}\Gamma^{i}
\mp\frac{1}{4}e^{\Phi}X_{i j}\Gamma^{i j} + e^{\Phi}m
\cr
&+& \Gamma_{11}\bigg(\pm \frac{1}{4}L_{i}\Gamma^{i} - \frac{1}{8}{\tilde{H}}_{i j k}\Gamma^{i j k} \mp e^{\Phi}S + \frac{1}{4}e^{\Phi}{\tilde{F}}_{i j}\Gamma^{i j} \bigg)~,
\eea
are the horizon Dirac operators associated with the supercovariant derivatives $\nabla^{(\pm)}$.

\vskip 0.3cm
{\it Theorem:}
Let ${\cal S}$ and the fields satisfy the conditions for the maximum principle to apply, e.g.~the fields are smooth and ${\cal S}$ is
compact without boundary. Then there is a 1-1 correspondence between the zero modes of ${\mathscr D}^{(+)}$ and the $\eta_+$ Killing spinors, i.e.
\bea
\nabla_{i}^{(+)}\eta_+=0~,~~~{\cal A}^{(+)}\eta_+=0~\Longleftrightarrow~ {\mathscr D}^{(+)}\eta_+=0~.
\eea
Moreover $\parallel\eta_+\parallel^2$ is constant.
\vskip 0.3cm

{\it Proof:} It is evident that if $\eta_+$ is a Killing spinor, then it is a zero mode of ${\mathscr D}^{(+)}$. To prove the converse, assuming that
$\eta_+$ is a zero mode of ${\mathscr D}^{(+)}$ and after using the field equations and Bianchi identities, one can establish the identity, see appendix C,
\bea
{\tilde{\nabla}}^{i}{\tilde{\nabla}}_{i}\parallel\eta_+\parallel^2 - (2\tilde{\nabla}^i \Phi +  h^i) {\tilde{\nabla}}_{i}\parallel\eta_+\parallel^2 = 2\parallel{\hat\nabla^{(+)}}\eta_{+}\parallel^2 + (-4\kappa - 16 \kappa^2)\parallel\mathcal{A}^{(+)}\eta_+\parallel^2~,
\label{maxprin}
\eea
where
\bea
\label{redef1}
\hat{\nabla}_{i}^{(\pm)}=\nabla_{i}^{(\pm)}+ \kappa \Gamma_i {\cal A}^{(\pm)}~,
\eea
for some $\kappa\in \bR$. Provided that $\kappa$ is chosen in the interval $(-{1\over4}, 0)$, the theorem follows as an application of the maximum
principle.

 \rightline{ $\square$}

Let us turn to investigate the relation between Killing spinors and the zero modes of the ${\mathscr D}^{(-)}$ operator.

\vskip 0.3cm
{\it Theorem:} Let ${\cal S}$ be compact without boundary and the horizon fields be smooth. There is a 1-1 correspondence between the
zero modes of ${\mathscr D}^{(-)}$ and the $\eta_-$ Killing spinors, i.e.
\bea
\nabla_{i}^{(-)}\eta_-=0~,~~~{\cal A}^{(-)}\eta_-=0~\Longleftrightarrow~ {\mathscr D}^{(-)}\eta_-=0~ \ .
\eea
\vskip 0.3cm

{\it Proof:} It is clear that if $\eta_-$ is a Killing spinor, then it is a zero mode of ${\mathscr D}^{(-)}$.  To prove the converse, if $\eta_-$
is a zero mode of ${\mathscr D}^{(-)}$, then upon using the field equations and Bianchi identities one can establish the formula, see appendix C,
\bea
\label{l2b}
{\tilde{\nabla}}^{i} \big( e^{-2 \Phi} V_i \big)
= -2 e^{-2 \Phi} \parallel{\hat\nabla^{(-)}}\eta_{-}\parallel^2 +   e^{-2 \Phi} (4 \kappa +16 \kappa^2) \parallel\mathcal{A}^{(-)}\eta_-\parallel^2~,
\nonumber \\
\eea
where $V=-d \parallel \eta_- \parallel^2 - \parallel \eta_- \parallel^2 h $. The theorem follows after integrating the above formula over ${\cal S}$ using Stokes' theorem
for  $\kappa\in (-{1\over4}, 0)$.

\rightline{ $\square$}

\subsection{Index theory and supersymmetry enhancement}
To prove the first part of the conjecture, we shall establish the theorem:

\vskip 0.3cm
{\it Theorem:} The number of supersymmetries preserved by massive IIA horizons is even.
\vskip 0.3cm

{\it Proof:} Let $N_\pm$ be the number of $\eta_\pm$ Killing spinors. As a consequence of the two theorems we have established in the previous section
$N_\pm=\mathrm{dim}\,\mathrm{Ker}\, {\mathscr D}^{(\pm)}$. The $Spin(9,1)$ bundle over the spacetime decomposes as  $S_+\oplus S_-$
upon restriction to ${\cal S}$. Furthermore $S_+$ and $S_-$ are isomorphic as $Spin(8)$ bundles as both are associated with the
Majorana representation.  The action of  ${\mathscr D}^{(+)}: \Gamma(S_+)\rightarrow \Gamma(S_+)$ on the section $ \Gamma(S_+)$  of $S_+$
is not  chirality preserving.  Since the principal symbol of ${\mathscr D}^{(+)}$ is the same as the principal symbol
of the standard Dirac operator acting on Majorana but not-Weyl spinors, the index vanishes \cite{index}.  Therefore
\bea
N_+=\mathrm{dim}\,\mathrm{Ker}\, {\mathscr D}^{(+)}= \mathrm{dim}\,\mathrm{Ker}\, ({\mathscr D}^{(+)})^\dagger~,
\eea
where $({\mathscr D}^{(+)})^\dagger$ is the adjoint of ${\mathscr D}^{(+)}$.  On the other hand, one can establish
\bea
\big(e^{2 \Phi} \Gamma_-\big) \big({\mathscr D}^{(+)}\big)^\dagger
= {\mathscr D}^{(-)} \big(e^{2 \Phi} \Gamma_-\big)~,
\eea
and so
\bea
N_-=\mathrm{dim}\,\mathrm{Ker}\, ({\mathscr D}^{(-)})=\mathrm{dim}\,\mathrm{Ker}\, ({\mathscr D}^{(+)})^\dagger~.
\eea
Therefore, we conclude that $N_+=N_-$ and so the number of supersymmetries of massive IIA horizons $N=N_++N_-=2 N_-$ is even.

\rightline{ $\square$}

\section{The $\mathfrak{sl}(2,\bR)$ symmetry of massive IIA horizons}

\subsection{ $\eta_+$ from $\eta_{-}$ Killing spinors}

We shall demonstrate the existence of the $\mathfrak{sl}(2,\bR)$ symmetry of massive IIA horizons by  directly constructing
the vector fields on the spacetime generated by the action of $\mathfrak{sl}(2,\bR)$. In turn the existence of such vector fields
is a consequence of the property that massive IIA horizons admit an even number of supersymmetries. We have seen that if $\eta_-$ is a Killing spinor, then  $\eta_+=\Gamma_+\Theta_-\eta_-$
is also a Killing spinor provided that $\eta_+\not=0$. It turns out that under certain conditions this is always possible.

\vskip 0.3cm
{\it Lemma:} Suppose that ${\cal S}$ and the fields satisfy the requirements for the maximum principle to apply. Then
\bea
\mathrm{Ker}\, \Theta_-=\{0\}~.
\label{kerz}
\eea
\vskip 0.3cm
{\it Proof:}
We shall prove this by contradiction. Assume that $\Theta_-$ has a non-trivial kernel, so there is $\eta_-\not=0$
such that $\Theta_- \eta_-=0$. In such a case,
  ({\ref{int3}}) gives
$ \Delta \langle \eta_- , \eta_- \rangle =0$. Thus $\Delta =0$,
as $\eta_-$ is no-where vanishing.

Next the gravitino KSE $\nabla^{(-)}\eta_-=0$ together with $\langle \eta_{-}, \Gamma_{i}\Theta_{-}\eta_{-} \rangle=0$ imply that
\begin{eqnarray}
\label{nrm1a}
{\tilde{\nabla}}_i \parallel \eta_-\parallel^2 = - h_i  \parallel \eta_-\parallel^2~.
\end{eqnarray}
On taking the divergence of this expression, eliminating  ${\tilde{\nabla}}^i h_i$ upon using ({\ref{feq7}}),
and after setting $\Delta=0$, one finds
\begin{eqnarray}
\label{nrm1ab}
{\tilde{\nabla}}^i {\tilde{\nabla}}_i  \parallel \eta_-\parallel^2 &=& 2\tilde{\nabla}^{i}\Phi {\tilde{\nabla}}_i \parallel \eta_-\parallel^2
\cr
&+& \bigg(L^2+ \frac{1}{2}e^{2\Phi}S^2 + \frac{1}{4}e^{2\Phi}X^2 + \frac{1}{4}e^{2\Phi}\tilde{F}^2 + \frac{1}{48}e^{2\Phi}\tilde{G}^2 + \frac{1}{2}e^{2\Phi}m^2\bigg)  \parallel \eta_-\parallel^2~.
\nonumber \\
\end{eqnarray}
The maximum principle implies that $\parallel \eta_- \parallel^2$ is constant. However, the remainder of
({\ref{nrm1ab}}) can never vanish, due to the quadratic term in $m$. So there can be no solutions, with $m \neq 0$, such
that $\eta_- \neq 0$ is in the Kernel of $\Theta_-$, and so $\mathrm{Ker}\, \Theta_-=\{0\}$.

\rightline{ $\square$}

\subsection{$\mathfrak{sl}(2,\bR)$ symmetry}

Using $\eta_-$ and $\eta_+=\Gamma_+\Theta_-\eta_-$ and the formula (\ref{lightconesol}),
one can construct two linearly independent Killing spinors on the  spacetime as
\bea
\epsilon_1=\eta_-+u\eta_++ru \Gamma_-\Theta_+\eta_+~,~~~\epsilon_2=\eta_++r\Gamma_-\Theta_+\eta_+~.
\eea
It is known from the general theory of supersymmetric massive IIA backgrounds that for any Killing spinors $\zeta_1$ and $\zeta_2$ the dual vector field $K(\zeta_1, \zeta_2)$ of the 1-form
bilinear
\bea
\omega(\zeta_1, \zeta_2)=\langle(\Gamma_+-\Gamma_-) \zeta_1, \Gamma_a\zeta_2\rangle\, e^a~,
\label{1formbi}
\eea
is a Killing vector and leaves invariant all the other fields of the theory.
Evaluating, the vector field bilinears of the Killing spinors $\epsilon_1$ and $\epsilon_2$, we find that
\begin{eqnarray}
K_1(\epsilon_1, \epsilon_2)&=&-2u \parallel\eta_+\parallel^2 \partial_u+ 2r \parallel\eta_+\parallel^2 \partial_r+ \tilde V~,
\cr
K_2(\epsilon_2, \epsilon_2)&=&-2 \parallel\eta_+\parallel^2 \partial_u~,
\cr
K_3(\epsilon_1, \epsilon_1)&=&-2u^2 \parallel\eta_+\parallel^2 \partial_u +(2 \parallel\eta_-\parallel^2+ 4ru \parallel\eta_+\parallel^2)\partial_r+ 2u \tilde V~,
\label{kkk}
\end{eqnarray}
where we have set
\begin{eqnarray}
\label{vii}
\tilde V =  \langle \Gamma_+ \eta_- , \Gamma^i \eta_+ \rangle\, \tilde \partial_i~,
\end{eqnarray}
is a vector field on ${\cal S}$.
To derive the above expressions for the Killing vector fields, we have used the identities
\begin{eqnarray}
- \Delta\, \parallel\eta_+\parallel^2 +4  \parallel\Theta_+ \eta_+\parallel^2 =0~,~~~\langle \eta_+ , \Gamma_i \Theta_+ \eta_+ \rangle  =0~,
\end{eqnarray}
which follow from the first integrability condition in ({\ref{int1}}),  $\parallel\eta_+\parallel=\mathrm{const}$ and the KSEs of $\eta_+$.

\vskip 0.3cm
{\it Theorem:} The Lie bracket algebra of  $K_1$, $K_2$ and $K_3$  is $\mathfrak{sl}(2,\bR)$.
\vskip 0.3cm
{\it Proof:} Using the identities summarised in appendix D, one can demonstrate after a direct computation that
\begin{eqnarray}
[K_1,K_2]=2 \parallel\eta_+\parallel^2 K_2~,~~~[K_2, K_3]=-4 \parallel\eta_+\parallel^2 K_1  ~,~~~[K_3,K_1]=2 \parallel\eta_+\parallel^2 K_3~. \ \
\end{eqnarray}
This proves the theorem and the last part of the conjecture.

\rightline{ $\square$}

\vskip 1cm

\noindent{\bf Acknowledgements} \vskip 0.1cm
UG is supported by the Knut and Alice Wallenberg Foundation. GP is partially supported by the STFC grant ST/J002798/1. JG is supported by the STFC grant, ST/1004874/1.
JG would like to thank the
Department of Mathematical Sciences, University of Liverpool for hospitality during which part of this work
was completed. UK is supported by a STFC PhD fellowship.

\vskip 0.5cm


\setcounter{section}{0}
\setcounter{equation}{0}

\appendix{Horizon Field equations and Bianchi Identities}

The bosonic fields   of massive IIA supergravity \cite{ romans} are the spacetime metric  $g$, the dilaton $\Phi$, the 2-form NS-NS gauge potential $B$,
and the 1-form and the 3-form RR gauge potentials  $A$ and $C$, respectively. The theory also includes a mass parameter $m$ which induces a negative cosmological constant in the theory.
 In addition,
 fermionic fields of the theory are  a Majorana gravitino and   dilatino  which  are set to zero
in all the computations that follow.
The bosonic field strengths of massive IIA supergravity \cite{romans} in the conventions of \cite{roo} are
\bea
F=dA + mB ~,~~~H=dB~,~~~G=dC-H\wedge A + \frac{1}{2}m B \wedge B~,
\eea
implying the Bianchi identities
\bea
\label{bian1}
dF=mH~,~~~dH=0~,~~~dG=F\wedge H~.
\eea
The bosonic part of the massive IIA action in the string frame is
\bea
S = \int \bigg[d^{10}x\,\sqrt{-g} \bigg(e^{-2 \Phi} \big(R+4 \nabla_\mu \Phi\nabla^\mu \Phi -{1 \over 12} H_{\lambda_1 \lambda_2 \lambda_3}
H^{\lambda_1 \lambda_2 \lambda_3} \big)
\nonumber \\
-{1 \over 4} F_{\mu \nu} F^{\mu \nu}
-{1 \over 48} G_{\mu_1 \mu_2 \mu_3 \mu_4}
G^{\mu_1 \mu_2 \mu_3 \mu_4} - \frac{1}{2}m^2 \bigg)
\nonumber \\
+ {1 \over 2} dC \wedge dC \wedge B +{m \over 6} dC \wedge B \wedge B \wedge B
\nonumber \\
+{m^2 \over 40} B \wedge B \wedge B \wedge B \wedge B \bigg]  ~.
\eea
This leads to the Einstein equation
\bea
\label{eineq}
R_{\mu \nu}&=&-2 \nabla_\mu \nabla_\nu \Phi
+{1 \over 4} H_{\mu \lambda_1 \lambda_2} H_\nu{}^{\lambda_1 \lambda_2}
+{1 \over 2} e^{2 \Phi} F_{\mu \lambda} F_\nu{}^\lambda
+{1 \over 12} e^{2 \Phi} G_{\mu \lambda_1 \lambda_2 \lambda_3}
G_\nu{}^{\lambda_1 \lambda_2 \lambda_3}
\nonumber \\
&+& g_{\mu \nu} \bigg(-{1 \over 8}e^{2 \Phi}
F_{\lambda_1 \lambda_2}F^{\lambda_1 \lambda_2}
-{1 \over 96}e^{2 \Phi} G_{\lambda_1 \lambda_2 \lambda_3
\lambda_4} G^{\lambda_1 \lambda_2 \lambda_3
\lambda_4} - \frac{1}{4}e^{2\Phi}m^2 \bigg)~,
\eea
and the dilaton field equation
\bea
\label{dileq}
\nabla^\mu \nabla_\mu \Phi
&=& 2 \nabla_\lambda \Phi \nabla^\lambda \Phi
-{1 \over 12} H_{\lambda_1 \lambda_2 \lambda_3}
H^{\lambda_1 \lambda_2 \lambda_3}+{3 \over 8} e^{2 \Phi}
F_{\lambda_1 \lambda_2} F^{\lambda_1 \lambda_2}
\nonumber \\
&+&{1 \over 96} e^{2 \Phi} G_{\lambda_1 \lambda_2 \lambda_3
\lambda_4} G^{\lambda_1 \lambda_2 \lambda_3
\lambda_4} + \frac{5}{4}e^{2\Phi}m^2 ~,
\eea
the 2-form field equation
\bea
\label{geq1}
\nabla^\mu F_{\mu \nu} +{1 \over 6} H^{\lambda_1 \lambda_2 \lambda_3} G_{\lambda_1 \lambda_2 \lambda_3 \nu} =0~,
\eea
the 3-form field equation
\bea
\label{beq1}
\nabla_\lambda \bigg( e^{-2 \Phi} H^{\lambda \mu \nu}\bigg) - mF^{\mu \nu} -{1 \over 2} G^{\mu \nu \lambda_1 \lambda_2} F_{\lambda_1 \lambda_2}
+{1 \over 1152} \epsilon^{\mu \nu \lambda_1 \lambda_2
\lambda_3 \lambda_4 \lambda_5 \lambda_6 \lambda_7 \lambda_8}
G_{\lambda_1 \lambda_2 \lambda_3 \lambda_4}
G_{\lambda_5 \lambda_6 \lambda_7 \lambda_8} =0~,
\nonumber \\
\eea
and the 4-form field equation
\bea
\label{ceq1}
\nabla_\mu G^{\mu \nu_1 \nu_2 \nu_3}
+{1 \over 144} \epsilon^{\nu_1 \nu_2 \nu_3
\lambda_1 \lambda_2 \lambda_3 \lambda_4 \lambda_5
\lambda_6 \lambda_7} G_{\lambda_1 \lambda_2 \lambda_3
\lambda_4} H_{\lambda_5 \lambda_6 \lambda_7}=0~.
\eea

Adapting  Gaussian null coordinates \cite{isen, gnull} near massive IIA Killing horizons, one finds
\be
ds^2 &=&2 \bbe^+ \bbe^- + \delta_{ij} \bbe^i \bbe^j~,~~~G= \bbe^+ \wedge \bbe^- \wedge X +r \bbe^+ \wedge Y + \tilde G~,~~
\cr
H &=&\bbe^+ \wedge \bbe^- \wedge L+ r \bbe^+ \wedge M + \tilde H~,~~~F= \bbe^+ \wedge \bbe^- S + r \bbe^+ \wedge T+ \tilde F~,
\la{hormetrx}
\ee
where  $\Delta$ is a function, $h$, $L$ and $T$ are 1-forms, $X$, $M$ and $\tilde F$ are 2-forms,
$Y, \tilde H$ are 3-forms and $\tilde G$ is a 4-form on  the spatial horizon section ${\cal S}$. The dilaton $\Phi$ is also taken as a function
on ${\cal S}$.

Substituting the fields (\ref{hormetr}) into the Bianchi identities  of massive IIA supergravity, one finds that
\bea
\label{bian2}
M&=&d_h L~,~~~T=d_h S - mL~,~~~Y=d_h X-L\wedge \tilde F-S \tilde H~,~~~
\cr
d\tilde G&=&\tilde H\wedge \tilde F~,~~~d\tilde H=0, \, d\tilde F=m\tilde{H}~,
\eea
where  $d_h \theta \equiv d \theta-
h \wedge \theta$ for any form $\theta$.

Similarly, the independent field equations of the near horizon fields are as follows. The 2-form field equation ({\ref{geq1}) gives
\bea
\label{feq1}
\tn^i {\tilde{F}}_{ik}-h^i {\tilde{F}}_{ik}
+T_k -L^i X_{ik} +{1 \over 6} {\tilde{H}}^{\ell_1 \ell_2
\ell_3} {\tilde{G}}_{\ell_1 \ell_2 \ell_3 k}=0~,
\eea
the 3-form field equation  ({\ref{beq1}}) gives
\bea
\label{feq2}
\tn^i (e^{-2 \Phi}L_i)- mS -{1 \over 2} {\tilde{F}}^{ij}
X_{ij} +{1 \over 1152}
\epsilon^{\ell_1 \ell_2 \ell_3 \ell_4 \ell_5 \ell_6 \ell_7
\ell_8} {\tilde{G}}_{\ell_1 \ell_2 \ell_3 \ell_4}
{\tilde{G}}_{\ell_5 \ell_6 \ell_7 \ell_8}=0~,
\eea
and
\bea
\label{feq3}
\tn^i(e^{-2 \Phi} {\tilde{H}}_{imn})-m\tilde{F}_{m n}
-e^{-2 \Phi} h^i {\tilde{H}}_{imn}
+e^{-2 \Phi} M_{mn}+S X_{mn}-{1 \over 2}
{\tilde{F}}^{ij} {\tilde{G}}_{ijmn}
\nonumber \\
-{1 \over 48} \epsilon_{mn}{}^{\ell_1
\ell_2 \ell_3 \ell_4 \ell_5 \ell_6} X_{\ell_1 \ell_2}
{\tilde{G}}_{\ell_3 \ell_4 \ell_5 \ell_6} =0~,
\eea
and the 4-form field equation ({\ref{ceq1}}) gives
\bea
\label{feq4}
\tn^i X_{ik} +{1 \over 144} \epsilon_k{}^{\ell_1 \ell_2
\ell_3 \ell_4 \ell_5 \ell_6 \ell_7}
{\tilde{G}}_{\ell_1 \ell_2 \ell_3 \ell_4} {\tilde{H}}_{\ell_5 \ell_6 \ell_7} =0~,
\eea
and
\bea
\label{feq5}
\tn^i {\tilde{G}}_{ijkq}+Y_{jkq}-h^i {\tilde{G}}_{ijkq}
-{1 \over 12} \epsilon_{jkq}{}^{\ell_1 \ell_2 \ell_3
\ell_4 \ell_5} X_{\ell_1 \ell_2} {\tilde{H}}_{\ell_3 \ell_4 \ell_5}
-{1 \over 24}\epsilon_{jkq}{}^{\ell_1 \ell_2 \ell_3
\ell_4 \ell_5} {\tilde{G}}_{\ell_1 \ell_2 \ell_3 \ell_4}
L_{\ell_5} =0~,
\nonumber \\
\eea
where $\tilde \nabla$ is the Levi-Civita connection of the metric on ${\cal S}$.
In addition, the dilaton field equation ({\ref{dileq}}) becomes
\bea
\label{feq6}
\tn^i \tn_i \Phi - h^i \tn_i \Phi &=&
2 \tn_i \Phi \tn^i \Phi +{1 \over 2} L_i L^i
-{1 \over 12} {\tilde{H}}_{\ell_1 \ell_2 \ell_3}
{\tilde{H}}^{\ell_1 \ell_2 \ell_3}-{3 \over 4} e^{2 \Phi}S^2
\nonumber \\
&+&{3 \over 8} e^{2 \Phi} {\tilde{F}}_{ij}
{\tilde{F}}^{ij} -{1 \over 8} e^{2 \Phi} X_{ij}X^{ij}
+{1 \over 96} e^{2 \Phi} {\tilde{G}}_{\ell_1 \ell_2 \ell_3
\ell_4} {\tilde{G}}^{\ell_1 \ell_2 \ell_3 \ell_4} + \frac{5}{4}e^{2\Phi}m^2~.
\nonumber \\
\eea
It remains to evaluate the Einstein field equation. This gives
\bea
\label{feq7}
{1 \over 2} \tn^i h_i -\Delta -{1 \over 2}h^2
&=& h^i \tn_i \Phi -{1 \over 2} L_i L^i -{1 \over 4} e^{2 \Phi} S^2 -{1 \over 8} e^{2 \Phi} X_{ij} X^{ij}
\nonumber \\
&-&{1 \over 8} e^{2 \Phi} {\tilde{F}}_{ij} {\tilde{F}}^{ij}
-{1 \over 96} e^{2 \Phi} {\tilde{G}}_{\ell_1 \ell_2
\ell_3 \ell_4} {\tilde{G}}^{\ell_1 \ell_2 \ell_3 \ell_4} - \frac{1}{4}e^{2\Phi}m^2~,
\eea
and
\bea
\label{feq8}
{\tilde{R}}_{ij} &=& -\tn_{(i} h_{j)}
+{1 \over 2} h_i h_j -2 \tn_i \tn_j \Phi
-{1 \over 2} L_i L_j +{1 \over 4} {\tilde{H}}_{i
\ell_1 \ell_2} {\tilde{H}}_j{}^{\ell_1 \ell_2}
\nonumber \\
&+&{1 \over 2} e^{2 \Phi} {\tilde{F}}_{i \ell}
{\tilde{F}}_j{}^\ell -{1 \over 2} e^{2 \Phi} X_{i \ell}
X_j{}^\ell+{1 \over 12} e^{2 \Phi} {\tilde{G}}_{i \ell_1 \ell_2 \ell_3} {\tilde{G}}_j{}^{\ell_1 \ell_2 \ell_3}
\nonumber \\
&+&\delta_{ij} \bigg({1 \over 4} e^{2 \Phi} S^2 - \frac{1}{4}e^{2\Phi}m^2
-{1 \over 8} e^{2 \Phi} {\tilde{F}}_{\ell_1 \ell_2}
{\tilde{F}}^{\ell_1 \ell_2} +{1 \over 8} e^{2 \Phi}
X_{\ell_1 \ell_2}X^{\ell_1 \ell_2}
-{1 \over 96} e^{2 \Phi} {\tilde{G}}_{\ell_1 \ell_2
\ell_3 \ell_4} {\tilde{G}}^{\ell_1 \ell_2 \ell_3 \ell_4}
\bigg)~,
\nonumber \\
\eea
where $\tilde R$ denotes the Ricci tensor of ${\cal S}$.

There are additional Bianchi identities and field equations which however are not independent of those we have stated above. We give these because they are useful
in many of the intermediate computations. In particular, we have the additional Bianchi identities
\bea
dT + S dh + dS \wedge h + m dL&=&0~,
\nonumber \\
dM+L \wedge dh -h \wedge dL &=&0~,
\nonumber \\
dY + dh \wedge X - h \wedge dX + h \wedge (S {\tilde{H}}
+{\tilde{F}} \wedge L)+T \wedge {\tilde{H}}+{\tilde{F}} \wedge M &=&0~.
\eea
There are also additional field equations given by
\bea
\label{auxeq1}
-\tn^i T_i + h^i T_i -{1 \over 2} dh^{ij} {\tilde{F}}_{ij}
-{1 \over 2} X_{ij}M^{ij} -{1 \over 6} Y_{ijk}{\tilde{H}}^{ijk}=0~,
\eea
\bea
\label{auxeq2}
- \tn^i (e^{-2 \Phi}M_{ik}) + e^{-2 \Phi} h^i M_{ik}
-{1 \over 2} e^{-2 \Phi} dh^{ij} {\tilde{H}}_{ijk}
-T^i X_{ik} -{1 \over 2} {\tilde{F}}^{ij} Y_{ijk}
\nonumber \\
-mT_k -{1 \over 144} \epsilon_k{}^{\ell_1
\ell_2 \ell_3 \ell_4 \ell_5 \ell_6 \ell_7}
Y_{\ell_1 \ell_2 \ell_3} {\tilde{G}}_{\ell_4 \ell_5 \ell_6 \ell_7} =0~,
\eea
\bea
\label{auxeq3}
- \tn^i Y_{imn}+h^i Y_{imn}-{1 \over 2} dh^{ij}
{\tilde{G}}_{ijmn}
+{1 \over 36} \epsilon_{mn}{}^{\ell_1 \ell_2 \ell_3
\ell_4 \ell_5 \ell_6} Y_{\ell_1 \ell_2 \ell_3}
{\tilde{H}}_{\ell_4 \ell_5 \ell_6}
\nonumber \\
+{1 \over 48} \epsilon_{mn}{}^{\ell_1 \ell_2 \ell_3
\ell_4 \ell_5 \ell_6} {\tilde{G}}_{\ell_1 \ell_2 \ell_3
\ell_4} M_{\ell_5 \ell_6}=0~,
\eea
corresponding to equations obtained from
the $+$ component of ({\ref{geq1}}),
the $k$ component of ({\ref{beq1}}) and
the $mn$ component of ({\ref{ceq1}}) respectively.
However, ({\ref{auxeq1}}), ({\ref{auxeq2}}) and
({\ref{auxeq3}}) are implied by ({\ref{feq1}})-
({\ref{feq5}}) together with the Bianchi identities
({\ref{bian2}}).

Note also that the $++$ and $+i$ components of the
Einstein equation, which are
\bea
\label{auxeq4}
{1 \over 2} \tn^i \tn_i \Delta -{3 \over 2} h^i \tn_i \Delta-{1 \over 2} \Delta \tn^i h_i
+ \Delta h^2 +{1 \over 4} dh_{ij} dh^{ij}
&=&(\tn^i \Delta - \Delta h^i)\tn_i \Phi +{1 \over 4} M_{ij}
M^{ij}
\nonumber \\
&+&{1 \over 2} e^{2 \Phi} T_i T^i
+{1 \over 12} e^{2 \Phi} Y_{ijk} Y^{ijk}
\nonumber \\
\eea
and
\bea
\label{auxeq5}
{1 \over 2} \tn^j dh_{ij}-dh_{ij} h^j - \tn_i \Delta + \Delta h_i
&=& dh_i{}^j \tn_j \Phi -{1 \over 2} M_i{}^j L_j
+{1 \over 4} M_{\ell_1 \ell_2} {\tilde{H}}_i{}^{\ell_1 \ell_2}-{1 \over 2} e^{2 \Phi} S T_i
\nonumber \\
&+&{1 \over 2} e^{2 \Phi} T^j {\tilde{F}}_{ij}
-{1 \over 4} e^{2 \Phi} Y_i{}^{\ell_1 \ell_2}
X_{\ell_1 \ell_2} +{1 \over 12} e^{2 \Phi}
Y_{\ell_1 \ell_2 \ell_3} {\tilde{G}}_i{}^{\ell_1 \ell_2
\ell_3}~,
\nonumber \\
\eea
are implied by ({\ref{feq6}}), ({\ref{feq7}}), ({\ref{feq8}}),
together with ({\ref{feq1}})-({\ref{feq5}}),
and the Bianchi identities ({\ref{bian2}}).

\newsection{Integrability conditions and KSEs} \label{indkse}

Substituting the solution (\ref{lightconesol}) of the KSEs along the light cone directions back into the gravitino KSE (\ref{GKSE}), and appropriately expanding in the $r$ and $u$ coordinates, we find that
for  the $\mu = \pm$ components, one obtains  the additional conditions
\bea
\label{int1}
&&\bigg({1\over2}\Delta - {1\over8}(dh)_{ij}\Gamma^{ij} + {1\over8}M_{ij}\Gamma_{11}\Gamma^{ij} + 2\big( {1\over 4} h_i \Gamma^{i} - {1\over 4} L_{i}\Gamma_{11}\Gamma^{i}
\cr
&&- {1\over 16} e^{\Phi}\Gamma_{11}(-2S + \tilde{F}_{i j}\Gamma^{i j}) - \frac{1}{8\cdot4!}e^{\Phi}(12X_{i j}\Gamma^{i j} - \tilde{G}_{i j k l}\Gamma^{i j k l}) + \frac{1}{8}e^{\Phi}m \big)\Theta_{+} \bigg)\phi_{+} = 0~,
\nonumber \\
\eea
\bea
\label{int2}
&&\bigg(\frac{1}{4}\Delta h_i \Gamma^{i} - \frac{1}{4}\partial_{i}\Delta \Gamma^{i} + \big(-\frac{1}{8}(dh)_{ij}\Gamma^{ij} - \frac{1}{8}M_{ij}\Gamma^{ij}\Gamma_{11} - \frac{1}{4}e^{\Phi}T_{i}\Gamma^{i}\Gamma_{11} + \frac{1}{24}e^{\Phi}Y_{i j k}\Gamma^{i j k} \big) \Theta_{+} \bigg) \phi_{+} = 0~,
\nonumber \\
\eea
\bea
\label{int3}
&&\bigg(-\frac{1}{2}\Delta - \frac{1}{8}(dh)_{ij}\Gamma^{ij} + \frac{1}{8}M_{ij}\Gamma^{ij}\Gamma_{11} - \frac{1}{4}e^{\Phi} T_{i}\Gamma^{i} \Gamma_{11} - \frac{1}{24}e^{\Phi}Y_{ijk}\Gamma^{ijk} + 2\big( -{1\over4} h_i\Gamma^i
\cr
&&-{1\over4} \Gamma_{11} L_i \Gamma^i+{1\over16} e^\phi \Gamma_{11} (2 S+\tilde F_{ij} \Gamma^{ij}) -{1\over8 \cdot 4!} e^\phi (12 X_{ij} \Gamma^{ij}
+\tilde G_{ijkl} \Gamma^{ijkl})-\frac{1}{8}e^{\Phi}m \big) \Theta_{-} \bigg)\phi_{-} = 0 \ .
\nonumber \\
\eea

Similarly the $\mu=i$ component of the gravitino KSEs gives
\bea
\label{int4}
&&\nabla^{(\pm)}_i\phi_\pm=0~~~~
\eea
and
\bea
\label{int5}
&&\tilde \nabla_i \tau_{+} + \bigg( -\frac{3}{4}h_i - \frac{1}{16}e^{\Phi}X_{l_1 l_2}\Gamma^{l_1 l_2}\Gamma_{i} - \frac{1}{8\cdot4!}e^{\Phi} \tilde G_{l_1\cdots l_4}\Gamma^{l_1 \cdots l_4}\Gamma_{i} - \frac{1}{8}e^{\Phi}m\Gamma_{i}
\cr
&&- \Gamma_{11}(\frac{1}{4}L_i + \frac{1}{8}\tilde{H}_{i j k}\Gamma^{j k} + \frac{1}{8}e^{\Phi} S \Gamma_{i} +    \frac{1}{16}e^{\Phi}\tilde{F}_{l_1 l_2}\Gamma^{l_1 l_2}\Gamma_{i})\bigg )\tau_{+}
\cr
&&+ \bigg(-\frac{1}{4}(dh)_{ij}\Gamma^{j} - \frac{1}{4}M_{ij}\Gamma^{j}\Gamma_{11} + \frac{1}{8}e^{\Phi}T_{j}\Gamma^{j}\Gamma_{i}\Gamma_{11} + \frac{1}{48}e^{\Phi}Y_{l_1 l_2 l_3}\Gamma^{l_1 l_2 l_3}\Gamma_{i} \bigg)\phi_{+} = 0~,
\nonumber \\
\eea
where we have set
\bea
\label{int6}
\tau_{+} = \Theta_{+}\phi_{+} \ .
\eea
We shall demonstrate that all the above conditions are not independent and follow  upon using the field equations and the Bianchi identities from those in (\ref{covr}).

Similarly,
substituting the solution of the KSEs (\ref{lightconesol})  into the dilatino KSE (\ref{AKSE}) and expanding appropriately in the $r$ and $u$ coordinates, we find
\bea
\label{int7}
&&\partial_i \Phi \Gamma^i \phi_{\pm} -{1\over12} \Gamma_{11} (\mp 6 L_i \Gamma^i+\tilde H_{ijk} \Gamma^{ijk}) \phi_\pm+{3\over8} e^\Phi \Gamma_{11} (\mp2 S+\tilde F_{ij} \Gamma^{ij})\phi_\pm
\cr
&&
+{1\over 4\cdot 4!}e^{\Phi} (\mp 12 X_{ij} \Gamma^{ij}+\tilde G_{j_1j_2j_3j_4} \Gamma^{j_1j_2j_3j_4}) \phi_\pm + \frac{5}{4}e^{\Phi}m\phi_\pm=0 \ ,
\eea
\be
\label{int8}
&&-\bigg( \partial_{i}\Phi\Gamma^{i} + \frac{1}{12}\Gamma_{11} (6L_i \Gamma^{i} + \tilde{H}_{ijk}\Gamma^{ijk}) + \frac{3}{8}e^{\Phi}\Gamma_{11}(2S + \tilde{F}_{ij}\Gamma^{ij})
\cr
&&- \frac{1}{4\cdot 4!}e^{\Phi}(12X_{ij}\Gamma^{ij} + \tilde{G}_{ijkl}\Gamma^{ijkl}) -  \frac{5}{4}e^{\Phi}m \bigg)\tau_{+}
\cr
&&+ \bigg(\frac{1}{4}M_{ij}\Gamma^{ij}\Gamma_{11} + \frac{3}{4}e^{\Phi}T_{i}\Gamma^{i}\Gamma_{11} + \frac{1}{24}e^{\Phi}Y_{ijk}\Gamma^{ijk}\bigg)\phi_{+}=0~.
\eea
Again, these are not  independent of those in (\ref{covr}).

\subsection{Independent KSEs}

\subsubsection{The (\ref{int5}) condition}

The (\ref{int5}) component of the KSEs is implied by (\ref{int4}), (\ref{int6}) and (\ref{int7}) together with a number of field equations and Bianchi identities. First evaluate the LHS of (\ref{int5}) by substituting in (\ref{int6}) to eliminate $\tau_+$, and use (\ref{int4}) to evaluate the supercovariant derivative of $\phi_+$. Also, using (\ref{int4}) one can compute
\bea
&&(\tilde{\nabla}_{j}\tilde{\nabla}_{i} - \tilde{\nabla}_{i}\tilde{\nabla}_{j})\phi_{+}
 = {1\over 4} \tilde \nabla_j (h_i) \phi_+ + {1\over4} \Gamma_{11} \tilde \nabla_j(L_i) \phi_+ -{1\over8} \Gamma_{11} \tilde \nabla_j (\tilde H_{i l_1 l_2}) \Gamma^{l_1 l_2} \phi_+
\cr
&&+{1\over16} e^\Phi \Gamma_{11} (- 2\tilde \nabla_j (S)+ \tilde \nabla_j (\tilde F_{kl}) \Gamma^{kl}) \Gamma_i \phi_+ - {1\over 8\cdot 4!} e^\Phi (- 12 \tilde \nabla_j (X_{kl}) \Gamma^{kl}+ \tilde \nabla_j (\tilde G_{j_1j_2j_3j_4}) \Gamma^{j_1j_2j_3j_4}) \Gamma_i \phi_+
\cr
&&+{1\over16} \tilde \nabla_j \Phi e^\Phi \Gamma_{11} (- 2 S+  \tilde F_{kl} \Gamma^{kl}) \Gamma_i \phi_+ - {1\over 8\cdot 4!} \tilde \nabla_j \Phi e^\Phi (- 12  X_{kl} \Gamma^{kl}+  \tilde G_{j_1j_2j_3j_4} \Gamma^{j_1j_2j_3j_4}) \Gamma_i \phi_+
\cr
&&- \frac{1}{8}e^{\Phi}\tilde{\nabla}_j \Phi m\Gamma_i \phi_+ + \big( {1\over 4} h_i  + {1\over4} \Gamma_{11} L_i  -{1\over8} \Gamma_{11} \tilde H_{ijk} \Gamma^{jk} +{1\over16} e^\Phi \Gamma_{11} (- 2 S+ \tilde F_{kl} \Gamma^{kl}) \Gamma_i
\cr
&&-{1\over 8\cdot 4!} e^\Phi (- 12 X_{kl} \Gamma^{kl}+ \tilde G_{j_1j_2j_3j_4} \Gamma^{j_1j_2j_3j_4}) \Gamma_i- \frac{1}{8}e^{\Phi} m\Gamma_i\big)\tilde \nabla_{j} \phi_+ - (i \leftrightarrow j) \ .
\label{DDphicond}
\eea
Then consider the following, where the first terms cancel from the definition of curvature,
\bea
\bigg(\frac{1}{4}\tilde{R}_{ij}\Gamma^{j} - \frac{1}{2}\Gamma^{j}(\tilde{\nabla}_{j}\tilde{\nabla}_{i} - \tilde{\nabla}_{i}\tilde{\nabla}_{j}) \bigg)\phi_+ + \frac{1}{2}\tilde{\nabla}_{i}(\mathcal{A}_1) + \frac{1}{2}\Psi_{i} \mathcal{A}_1 = 0~,
\label{B5intcond}
\eea
where
\bea
\mathcal{A}_1 &=& \partial_i \Phi \Gamma^i \phi_{+} -{1\over12} \Gamma_{11} (- 6 L_i \Gamma^i+\tilde H_{ijk} \Gamma^{ijk}) \phi_++{3\over8} e^\Phi \Gamma_{11} (-2 S+\tilde F_{ij} \Gamma^{ij})\phi_+
\cr
&
+&{1\over 4\cdot 4!}e^{\Phi} (- 12 X_{ij} \Gamma^{ij}+\tilde G_{j_1j_2j_3j_4} \Gamma^{j_1j_2j_3j_4}) \phi_+ +  \frac{5}{4}e^{\Phi} m\phi_+
\label{A1cond}
\eea
and
\bea
\Psi_{i} &=& - \frac{1}{4}h_{i} + \Gamma_{11}(\frac{1}{4}L_{i} - \frac{1}{8}\tilde{H}_{i j k}\Gamma^{j k})~.
\label{Psiicond}
\eea
The expression in (\ref{A1cond}) vanishes on making use of (\ref{int7}), as $\mathcal{A}_1 = 0$ is equivalent to the $+$ component of (\ref{int7}). However a non-trivial identity is obtained by using (\ref{DDphicond}) in (\ref{B5intcond}), and expanding out the $\mathcal{A}_1$ terms. Then, on adding (\ref{B5intcond}) to the LHS of (\ref{int5}), with $\tau_+$ eliminated in favour of $\eta_+$ as described above, one obtains the following
 \bea
&&\frac{1}{4}\bigg(\tilde{R}_{ij} + \tn_{(i} h_{j)}
-{1 \over 2} h_i h_j + 2 \tn_i \tn_j \Phi
+{1 \over 2} L_i L_j -{1 \over 4} {\tilde{H}}_{i
l_1 l_2} {\tilde{H}}_j{}^{l_1 l_2}
\cr
&-&{1 \over 2} e^{2 \Phi} {\tilde{F}}_{i l}
{\tilde{F}}_j{}^l + {1 \over 8} e^{2 \Phi} {\tilde{F}}_{l_1 l_2}
{\tilde{F}}^{l_1 l_2}\delta_{ij} + {1 \over 2} e^{2 \Phi} X_{i l}
X_j{}^l - {1 \over 8} e^{2 \Phi}
X_{l_1 l_2}X^{l_1 l_2}\delta_{ij}
\cr
&-&{1 \over 12} e^{2 \Phi} {\tilde{G}}_{i \ell_1 \ell_2 \ell_3} {\tilde{G}}_j{}^{\ell_1 \ell_2 \ell_3}
+ {1 \over 96} e^{2 \Phi} {\tilde{G}}_{\ell_1 \ell_2
\ell_3 \ell_4} {\tilde{G}}^{\ell_1 \ell_2 \ell_3 \ell_4}\delta_{ij} - {1 \over 4} e^{2 \Phi} S^2\delta_{ij} + {1 \over 4} e^{2 \Phi} m^2\delta_{ij} \bigg)\Gamma^{j}=0~.
\nonumber \\
\eea
This vanishes identically on making use of the Einstein equation (\ref{feq8}). Therefore it follows that (\ref{int5}) is implied by the $+$ component of (\ref{int4}), (\ref{int6}) and (\ref{int7}), the Bianchi identities (\ref{bian2}) and the gauge field equations (\ref{feq1})-(\ref{feq5}).

\subsubsection{The (\ref{int8}) condition}
Let us define
\bea
\mathcal{A}_2 =
&&-\bigg( \partial_{i}\Phi\Gamma^{i} + \frac{1}{12}\Gamma_{11} (6L_i \Gamma^{i} + \tilde{H}_{ijk}\Gamma^{ijk}) + \frac{3}{8}e^{\Phi}\Gamma_{11}(2S + \tilde{F}_{ij}\Gamma^{ij})
\cr
&&- \frac{1}{4\cdot 4!}e^{\Phi}(12X_{ij}\Gamma^{ij} + \tilde{G}_{ijkl}\Gamma^{ijkl}) - \frac{5}{4}e^{\Phi}m\bigg)\tau_{+}
\cr
&&+ \bigg(\frac{1}{4}M_{ij}\Gamma^{ij}\Gamma_{11} + \frac{3}{4}e^{\Phi}T_{i}\Gamma^{i}\Gamma_{11} + \frac{1}{24}e^{\Phi}Y_{ijk}\Gamma^{ijk}\bigg)\phi_{+}~,
\eea
where $\mathcal{A}_2$ equals the expression in (\ref{int8}).
One obtains the following identity
\bea
\mathcal{A}_2 = -\frac{1}{2}\Gamma^{i}\tilde{\nabla}_i \mathcal{A}_1 + \Psi_1\mathcal{A}_1 ~,
\eea
where
\bea
\Psi_1 &&= \tilde{\nabla}_{i}\Phi\Gamma^{i} + \frac{3}{8}h_{i}\Gamma^{i} + \frac{1}{16}e^{\Phi}X_{l_1 l_2}\Gamma^{l_1 l_2} - \frac{1}{192}e^{\Phi}\tilde{G}_{l_1 l_2 l_3 l_4}\Gamma^{l_1 l_2 l_3 l_4} - \frac{1}{8}e^{\Phi}m
\cr
&&+ \Gamma_{11}\bigg(\frac{1}{48}\tilde{H}_{l_1 l_2 l_3}\Gamma^{l_1 l_2 l_3} - \frac{1}{8}L_{i}\Gamma^{i} + \frac{1}{16}e^{\Phi}\tilde{F}_{l_1 l_2}\Gamma^{l_1 l_2} - \frac{1}{8}e^{\Phi}S \bigg)~.
\eea
We have made use of the $+$ component of (\ref{int4}) in order to evaluate the covariant derivative in the above expression. In addition we have made use of the Bianchi identities (\ref{bian2}) and the field equations (\ref{feq1})-(\ref{feq6}).

\subsubsection{The (\ref{int1}) condition}
\label{B1sec}

In order to show that (\ref{int1}) is implied by the independent KSEs we can compute the following,
\bea
&&\bigg(-\frac{1}{4}\tilde{R} - \Gamma^{i j}\tilde{\nabla}_{i}\tilde{\nabla}_{j}\bigg)\phi_{+} - \Gamma^{i}\tilde{\nabla}_i(\mathcal{A}_1)
\cr
&+& \bigg(\tilde{\nabla}_i\Phi \Gamma^{i} + \frac{1}{4}h_{i}\Gamma^{i} + \frac{1}{16}e^{\Phi}X_{l_1 l_2}\Gamma^{l_1 l_2} - \frac{1}{192}e^{\Phi}\tilde{G}_{l_1 l_2 l_3 l_4}\Gamma^{l_1 l_2 l_3 l_4} - \frac{1}{8}e^{\Phi}m
\cr
&+& \Gamma_{11}(-\frac{1}{4}L_{l}\Gamma^{l} - \frac{1}{24}\tilde{H}_{l_1 l_2 l_3}\Gamma^{l_1 l_2 l_3} - \frac{1}{8}e^{\Phi}S
+ \frac{1}{16}e^{\Phi}\tilde{F}_{l_1 l_2}\Gamma^{l_1 l_2}) \bigg)\mathcal{A}_1 = 0~,
\eea
where
\bea
\tilde{R} &&= -2\Delta - 2h^{i}\tilde{\nabla}_{i}\Phi - 2\tilde{\nabla}^2\Phi - \frac{1}{2}h^2 + \frac{1}{2}L^2 + \frac{1}{4}\tilde{H}^{2} + \frac{5}{2}e^{2\Phi}S^2
\cr
&&- \frac{1}{4}e^{2\Phi}\tilde{F}^2 + \frac{3}{4}e^{2\Phi}X^2 + \frac{1}{48}e^{2\Phi}\tilde{G}^2 - \frac{3}{2}e^{2\Phi}m^2
\eea
and where we use the $+$ component of (\ref{int4}) to evaluate the covariant derivative terms. In order to obtain (\ref{int1}) from these expressions we make use of the Bianchi identities (\ref{bian2}), the field equations (\ref{feq1})-(\ref{feq6}), in particular in order to eliminate the $(\tilde{\nabla} \Phi)^2$ term. We have also made use of the $+-$ component of the Einstein equation (\ref{feq7}) in order to rewrite the scalar curvature $\tilde{R}$ in terms of $\Delta$. Therefore (\ref{int1}) follows from (\ref{int4}) and (\ref{int7}) together with the field equations and Bianchi identities mentioned above.

\subsubsection{The + (\ref{int7}) condition linear in $u$}
Since $\phi_+ = \eta_+ + u\Gamma_{+}\Theta_{-}\eta_-$, we must consider the part of the $+$ component of (\ref{int7}) which is linear in $u$. On defining
\bea
\mathcal{B}_1 &=& \partial_i \Phi \Gamma^i \eta_{-} -{1\over12} \Gamma_{11} ( 6 L_i \Gamma^i+\tilde H_{ijk} \Gamma^{ijk}) \eta_-+{3\over8} e^\Phi \Gamma_{11} (2 S+\tilde F_{ij} \Gamma^{ij})\eta_-
\cr
&&
+{1\over 4\cdot 4!}e^{\Phi} ( 12 X_{ij} \Gamma^{ij}+\tilde G_{j_1j_2j_3j_4} \Gamma^{j_1j_2j_3j_4}) \eta_- + \frac{5}{4}e^{\Phi}m ~ \eta_-
\eea
one finds that the $u$-dependent part of (\ref{int7}) is proportional to
\bea
-\frac{1}{2}\Gamma^{i}\tilde{\nabla}_{i}(\mathcal{B}_1) + \Psi_2 \mathcal{B}_1~,
\eea
where
\bea
\Psi_2 &&= \tilde{\nabla}_{i}\Phi\Gamma^{i} + \frac{1}{8}h_{i}\Gamma^{i} - \frac{1}{16}e^{\Phi}X_{l_1 l_2}\Gamma^{l_1 l_2} - \frac{1}{192}e^{\Phi}\tilde{G}_{l_1 l_2 l_3 l_4}\Gamma^{l_1 l_2 l_3 l_4} - \frac{1}{8}e^{\Phi}m
\cr
&&+ \Gamma_{11}\bigg(\frac{1}{48}\tilde{H}_{l_1 l_2 l_3}\Gamma^{l_1 l_2 l_3} + \frac{1}{8}L_{i}\Gamma^{i} + \frac{1}{16}e^{\Phi}\tilde{F}_{l_1 l_2}\Gamma^{l_1 l_2} + \frac{1}{8}e^{\Phi}S \bigg)~.
\eea
We have made use of the $-$ component of (\ref{int4}) in order to evaluate the covariant derivative in the above expression. In addition we have made use of the Bianchi identities (\ref{bian2}) and the field equations (\ref{feq1})-(\ref{feq6}).

\subsubsection{The (\ref{int2}) condition }
In order to show that (\ref{int2}) is implied by the independent KSEs we will show that it follows from (\ref{int1}). First act on (\ref{int1}) with the Dirac operator $\Gamma^{i}\tilde{\nabla}_{i}$ and use the field equations (\ref{feq1}) - (\ref{feq6}) and the Bianchi identities to eliminate the terms which contain derivatives of the fluxes and then use (\ref{int1}) to rewrite the $dh$-terms in terms of $\Delta$. Then use the conditions (\ref{int4}) and (\ref{int5}) to eliminate the $\partial_i \Phi$-terms from the resulting expression, some of the remaining terms will vanish as a consequence of (\ref{int1}). After performing these calculations, the condition (\ref{int2}) is obtained, therefore it follows from section \ref{B1sec} above that (\ref{int2}) is implied by (\ref{int4}) and (\ref{int7}) together with the field equations and Bianchi identities mentioned above.

\subsubsection{The (\ref{int3}) condition }
In order to show that (\ref{int3}) is implied by the independent KSEs we can compute the following,
\bea
&&\bigg(\frac{1}{4}\tilde{R} + \Gamma^{i j}\tilde{\nabla}_{i}\tilde{\nabla}_{j}\bigg)\eta_- + \Gamma^{i}\tilde{\nabla}_i(\mathcal{B}_1)
\cr
&+& \bigg(-\tilde{\nabla}_i\Phi \Gamma^{i} + \frac{1}{4}h_{i}\Gamma^{i} + \frac{1}{16}e^{\Phi}X_{l_1 l_2}\Gamma^{l_1 l_2} + \frac{1}{192}e^{\Phi}\tilde{G}_{l_1 l_2 l_3 l_4}\Gamma^{l_1 l_2 l_3 l_4} + \frac{1}{8}e^{\Phi}m
\cr
&+& \Gamma_{11}(-\frac{1}{4}L_{l}\Gamma^{l} + \frac{1}{24}\tilde{H}_{l_1 l_2 l_3}\Gamma^{l_1 l_2 l_3} - \frac{1}{8}e^{\Phi}S
- \frac{1}{16}e^{\Phi}\tilde{F}_{l_1 l_2}\Gamma^{l_1 l_2}) \bigg)\mathcal{B}_1 = 0~,
\eea
where we use the $-$ component of (\ref{int4}) to evaluate the covariant derivative terms. The expression above vanishes identically since the $-$ component of (\ref{int7}) is equivalent to $\mathcal{B}_1 = 0$. In order to obtain (\ref{int3}) from these expressions we make use of the Bianchi identities (\ref{bian2}) and the field equations (\ref{feq1})-(\ref{feq6}). Therefore (\ref{int3}) follows from (\ref{int4}) and (\ref{int7}) together with the field equations and Bianchi identities mentioned above.

\subsubsection{The + (\ref{int4}) condition linear in $u$}

Next consider the part of the $+$ component of (\ref{int4}) which is linear in $u$. First compute
\bea
\bigg(\Gamma^{j}(\tilde{\nabla}_{j}\tilde{\nabla}_{i} - \tilde{\nabla}_{i}\tilde{\nabla}_{j})  -\frac{1}{2}\tilde{R}_{ij}\Gamma^{j}\bigg)\eta_- - \tilde{\nabla}_{i}(\mathcal{B}_1) - \Psi_{i} \mathcal{B}_1 = 0~,
\eea
where
\bea
\Psi_{i} &=& \frac{1}{4}h_{i} - \Gamma_{11}(\frac{1}{4}L_{i} + \frac{1}{8}\tilde{H}_{i j k}\Gamma^{j k})
\eea
and where we have made use of the $-$ component of (\ref{int4}) to evaluate the covariant derivative terms. The resulting expression corresponds to the expression obtained by expanding out the $u$-dependent part of the $+$ component of (\ref{int4}) by using the $-$ component of (\ref{int4}) to evaluate the covariant derivative. We have made use of the Bianchi identities (\ref{bian2}) and the field equations (\ref{feq1})-(\ref{feq5}).


\appendix{Calculation of Laplacian of $\parallel \eta_\pm \parallel^2$}
\label{maxpex}

To establish the Lichnerowicz type theorems in \ref{lichthx}, we calculate the Laplacian of $\parallel \eta_\pm \parallel^2$.
For this let us generalise the modified horizon Dirac operator as ${\mathscr D}^{(\pm)}= {\cal D}^{(\pm)}+ q {\cal A}^{(\pm)}$ and assume throughout that  ${\mathscr D}^{(\pm)}\eta_\pm=0$;
in section \ref{lichthx} we had set $q=-1$.

To proceed, we compute the Laplacian
\bea
\tilde{\nabla}^i \tilde{\nabla}_i ||\eta_{\pm}||^2 = 2\langle\eta_\pm,\tilde{\nabla}^i \tilde{\nabla}_i\eta_\pm\rangle + 2 \langle\tilde{\nabla}^i \eta_\pm, \tilde{\nabla}_i \eta_\pm\rangle \ .
\eea
To evaluate this expression note that
\bea
\tilde{\nabla}^i \tilde{\nabla}_i \eta_\pm &=& \Gamma^{i}\tilde{\nabla}_{i}(\Gamma^{j}\tilde{\nabla}_j \eta_\pm) -\Gamma^{i j}\tilde{\nabla}_i \tilde{\nabla}_j \eta_\pm
\nonumber \\
&=& \Gamma^{i}\tilde{\nabla}_{i}(\Gamma^{j}\tilde{\nabla}_j \eta_\pm) + \frac{1}{4}\tilde{R}\eta_\pm
\nonumber \\
&=& \Gamma^{i}\tilde{\nabla}_{i}(-\Psi^{(\pm)}\eta_\pm -q\mathcal{A}^{(\pm)}\eta_{\pm}) + \frac{1}{4}\tilde{R} \eta_\pm \ .
\eea
It follows that
\bea
\langle\eta_\pm,\tilde{\nabla}^i \tilde{\nabla}_i\eta_\pm \rangle &=& \frac{1}{4}\tilde{R}\parallel \eta_\pm \parallel^2
+ \langle\eta_\pm, \Gamma^{i}\tilde{\nabla}_i(-\Psi^{(\pm)} - q\mathcal{A}^{(\pm)})\eta_\pm\rangle
\nonumber \\
&+& \langle\eta_\pm, \Gamma^{i}(-\Psi^{(\pm)} - q\mathcal{A}^{(\pm)})\tilde{\nabla}_i \eta_\pm \rangle~,
\eea
and also
\bea
\langle\tilde{\nabla}^i \eta_\pm, \tilde{\nabla}_i \eta_\pm\rangle &=& \langle{\hat\nabla^{(\pm)i}} \eta_\pm, {\hat\nabla^{(\pm)}_{i}} \eta_\pm\rangle - 2\langle\eta_\pm, (\Psi^{(\pm)i} + \kappa\Gamma^{i}\mathcal{A}^{(\pm)})^{\dagger} \tilde{\nabla}_i \eta_\pm \rangle
\nonumber \\
&-& \langle\eta_\pm, (\Psi^{(\pm)i} + \kappa\Gamma^{i}\mathcal{A}^{(\pm)})^{\dagger} (\Psi^{(\pm)}_i + \kappa \Gamma_{i} \, \mathcal{A}^{(\pm)}) \eta_\pm \rangle
\nonumber \\
&=& \parallel {\hat\nabla^{(\pm)}}\eta_{\pm} \parallel^2 - 2\langle \eta_{\pm}, \Psi^{(\pm)i\dagger}\tilde{\nabla}_{i}\eta_{\pm}\rangle
- 2\kappa \langle \eta_{\pm}, \mathcal{A}^{(\pm)\dagger}\Gamma^{i}\tilde{\nabla}_{i}\eta_{\pm}\rangle
\nonumber \\
&-&  \langle \eta_\pm, (\Psi^{(\pm)i\dagger}\Psi^{(\pm)}_i + 2\kappa \mathcal{A}^{(\pm)\dagger}\Psi^{(\pm)} + 8\kappa^2\mathcal{A}^{(\pm)\dagger}\mathcal{A}^{(\pm)})\eta_\pm \rangle
\nonumber \\
&=& \parallel {\hat\nabla^{(\pm)}}\eta_{\pm} \parallel^2 - 2\langle \eta_{\pm}, \Psi^{(\pm)i\dagger}\tilde{\nabla}_{i}\eta_{\pm}\rangle - \langle \eta_\pm , \Psi^{(\pm)i\dagger}\Psi^{(\pm)}_i \eta_\pm \rangle
\nonumber \\
&+& (2\kappa q - 8\kappa^2)\parallel \mathcal{A}^{(\pm)}\eta_\pm \parallel^2 \ .
\eea
Therefore,
\bea
\label{extralap1b}
\frac{1}{2}\tilde{\nabla}^i \tilde{\nabla}_i ||\eta_{\pm}||^2 &=& \parallel {\hat\nabla^{(\pm)}}\eta_{\pm} \parallel^2 + \, (2\kappa q - 8\kappa^2)\parallel \mathcal{A}^{(\pm)}\eta_\pm \parallel^2
\nonumber \\
&+& \langle \eta_\pm, \bigg(\frac{1}{4}\tilde{R} + \Gamma^{i}\tilde{\nabla}_i(-\Psi^{(\pm)} - q\mathcal{A}^{(\pm)}) - \Psi^{(\pm)i\dagger}\Psi^{(\pm)}_i \bigg) \eta_\pm \rangle
\nonumber \\
&+& \langle \eta_\pm, \bigg( \Gamma^{i}(-\Psi^{(\pm)} - q\mathcal{A}^{(\pm)}) - 2\Psi^{(\pm)i\dagger}\bigg)\tilde{\nabla}_i \eta_\pm \rangle \ .
\eea
In order to simplify the expression for the Laplacian, we shall attempt to rewrite the third line in ({\ref{extralap1b}}) as
\bea
\label{bilin}
\langle \eta_\pm, \bigg( \Gamma^{i}(-\Psi^{(\pm)} - q\mathcal{A}^{(\pm)}) - 2\Psi^{(\pm)i\dagger}\bigg)\tilde{\nabla}_i \eta_\pm \rangle = \langle \eta_\pm, \mathcal{F}^{(\pm)}\Gamma^{i}\tilde{\nabla}_i \eta_\pm \rangle + W^{(\pm)i}\tilde{\nabla}_i \parallel \eta_\pm \parallel^2~,
\nonumber \\
\eea
where $\mathcal{F}^{(\pm)}$ is linear in the fields and $W^{(\pm)i}$ is a vector. This expression is particularly advantageous, because the
first term on the RHS can be rewritten using the horizon
Dirac equation, and the second term is consistent with the application
of the maximum principle/integration by parts arguments which
are required for the generalised Lichnerowicz theorems. In order to rewrite ({\ref{bilin}}) in this fashion, note that
\bea
\Gamma^{i}(\Psi^{(\pm)} + q\mathcal{A}^{(\pm)}) + 2\Psi^{(\pm)i\dagger} &=& \big(\mp h^i \mp (q+1)\Gamma_{11}L^{i} + {1 \over 2}(q+1)\Gamma_{11}\tilde{H}^{i}{}_{\ell_1 \ell_2}\Gamma^{\ell_1 \ell_2}
+ 2q\tilde{\nabla}^i \Phi \big)
\nonumber \\
&+& \big(\pm \frac{1}{4}h_{j}\Gamma^{j} \pm (\frac{q}{2} + \frac{1}{4})\Gamma_{11} L_{j}\Gamma^{j}
\nonumber \\
&-& (\frac{q}{12} + \frac{1}{8})\Gamma_{11}\tilde{H}_{\ell_1 \ell_2 \ell_3}\Gamma^{\ell_1 \ell_2 \ell_3}- q\tilde{\nabla}_j \Phi \Gamma^{j}\big)\Gamma^{i}
\nonumber \\
&+& (q+1) \bigg(\mp {1 \over 8}e^{\Phi}X_{\ell_1 \ell_2}\Gamma^{i}\Gamma^{\ell_1 \ell_2} +{1 \over 96} e^{\Phi}\tilde{G}_{\ell_1 \ell_2 \ell_3 \ell_4}\Gamma^{i}\Gamma^{\ell_1 \ell_2 \ell_3 \ell_4} + \frac{5}{4}e^{\Phi}m\Gamma^i \bigg)
\nonumber \\
&+& (q+1)\Gamma_{11}\bigg(\pm {3 \over 4} e^{\Phi}S\Gamma^{i}
-{3 \over 8}e^{\Phi}\tilde{F}_{\ell_1 \ell_2}\Gamma^{i}\Gamma^{\ell_1 \ell_2}\bigg) \ .
\eea
One finds that (\ref{bilin}) is only
possible for $q=-1$ and thus we have
\bea
W^{(\pm)i} = \frac{1}{2}(2\tilde{\nabla}^i \Phi \pm h^i)
\eea
\bea
\mathcal{F}^{(\pm)} = \mp \frac{1}{4}h_{j}\Gamma^{j} - \tilde{\nabla}_{j}\Phi \Gamma^{j} + \Gamma_{11}\bigg(\pm \frac{1}{4}L_{j}\Gamma^{j} +  \frac{1}{24}\tilde{H}_{\ell_1 \ell_2 \ell_3}\Gamma^{\ell_1 \ell_2 \ell_3}\bigg) \ .
\eea

We remark that  $\dagger$ is the adjoint with respect to the $Spin(8)$-invariant inner product $\langle \phantom{i},\phantom{i} \rangle$. The choice of inner product is such that
\bea
\label{hermiden1}
\langle \eta_+, \Gamma^{[k]} \eta_+ \rangle &=& 0, \qquad
k = 2\, (\text{mod }4) \ {\rm and} \  k=3\, (\text{mod }4)
\cr
\langle \eta_+, \Gamma_{11}\Gamma^{[k]} \eta_+ \rangle &=& 0,
\qquad k = 1\, (\text{mod }4) \ {\rm and} \  k= 2\, (\text{mod }4) \ ,
\eea
where $\Gamma^{[k]}$ denote skew-symmetric products of k gamma matrices. For a more detailed explanation see \cite{iiaindex}.

It follows that
\bea
\label{laplacian}
\frac{1}{2}\tilde{\nabla}^i \tilde{\nabla}_i ||\eta_{\pm}||^2 &=& \parallel {\hat\nabla^{(\pm)}}\eta_{\pm} \parallel^2 + \, (-2\kappa  - 8\kappa^2)\parallel \mathcal{A}^{(\pm)}\eta_\pm \parallel^2
+ W^{(\pm)i}\tilde{\nabla}_{i}\parallel \eta_\pm \parallel^2
\nonumber \\
&+& \langle \eta_\pm, \bigg(\frac{1}{4}\tilde{R} + \Gamma^{i}\tilde{\nabla}_i(-\Psi^{(\pm)} + \mathcal{A}^{(\pm)}) - \Psi^{(\pm)i\dagger}\Psi^{(\pm)}_i  + \mathcal{F}^{(\pm)}(-\Psi^{(\pm)} + \mathcal{A}^{(\pm)})\bigg) \eta_\pm \rangle \ .
\nonumber \\
\eea
Using (\ref{feq8}) and the dilaton field equation (\ref{feq6}),  we get
\bea
\tilde{R} &=& -\tilde{\nabla}^{i}(h_i) + \frac{1}{2}h^2 - 4(\tilde{\nabla}\Phi)^2 - 2h^{i}\tilde{\nabla}_{i}\Phi - \frac{3}{2}L^2 + \frac{5}{12}\tilde{H}^2
\nonumber \\
&+& \frac{7}{2}e^{2\Phi}S^2 - \frac{5}{4}e^{2\Phi}\tilde{F}^2 + \frac{3}{4}e^{2\Phi}X^2 - \frac{1}{48}e^{2\Phi}\tilde{G}^2 - \frac{9}{2}e^{2\Phi}m^2 \ .
\eea
One obtains, upon using the field equations and Bianchi identities,
\bea
\label{quad}
\bigg(\frac{1}{4}\tilde{R} &+& \Gamma^{i}\tilde{\nabla}_i(-\Psi^{(\pm)} + \mathcal{A}^{(\pm)}) - \Psi^{(\pm)i\dagger}\Psi^{(\pm)}_i  + \mathcal{F}^{(\pm)}(-\Psi^{(\pm)} + \mathcal{A}^{(\pm)})\bigg)\eta_\pm
\nonumber \\
&=& \bigg[ \big(\pm \frac{1}{4}\tilde{\nabla}_{\ell_1}(h_{\ell_2}) \mp \frac{1}{16}\tilde{H}^{i}{}_{\ell_1 \ell_2}L_{i}\big)\Gamma^{\ell_1 \ell_2}+  \big( \pm \frac{1}{8}\tilde{\nabla}_{\ell_1}(e^{\Phi}X_{\ell_2 \ell_3}) + \frac{1}{24}\tilde{\nabla}^{i}(e^{\Phi}\tilde{G}_{i \ell_1 \ell_2 \ell_3})
\nonumber \\
&\mp& \frac{1}{96}e^{\Phi}h^{i}\tilde{G}_{i \ell_1 \ell_2 \ell_3} -\frac{1}{32}e^{\Phi}X_{\ell_1 \ell_2}h_{\ell_3}  \mp \frac{1}{8}e^{\Phi}\tilde{\nabla}_{\ell_1}\Phi X_{\ell_2 \ell_3}
- \frac{1}{24}e^{\Phi}\tilde{\nabla}^{i}\Phi \tilde{G}_{i \ell_1 \ell_2 \ell_3}
\nonumber \\
&\mp& \frac{1}{32}e^{\Phi}\tilde{F}_{\ell_1 \ell_2}L_{\ell_3}
\mp \frac{1}{96}e^{\Phi}S\tilde{H}_{\ell_1 \ell_2 \ell_3} - \frac{1}{32}e^{\Phi}\tilde{F}^{i}{}_{\ell_1}\tilde{H}_{i \ell_2 \ell_3}\big)\Gamma^{\ell_1 \ell_2 \ell_3}
\nonumber \\
&+& \Gamma_{11}\bigg(\big(\mp \frac{1}{4}\tilde{\nabla}_{\ell}(e^{\Phi}S) - \frac{1}{4}\tilde{\nabla}^{i}(e^{\Phi}\tilde{F}_{i \ell}) +\frac{1}{16}e^{\Phi}S h_{\ell} \pm \frac{1}{16}e^{\Phi}h^{i}\tilde{F}_{i \ell} \pm \frac{1}{4}e^{\Phi}\tilde{\nabla}_{\ell}\Phi S
\nonumber \\
&+& \frac{1}{4}e^{\Phi}\tilde{\nabla}^{i}\Phi\tilde{F}_{i \ell} + \frac{1}{16}e^{\Phi}L^i X_{i \ell}
\mp \frac{1}{32}e^{\Phi}\tilde{H}^{i j}{}_{\ell}X_{i j}
- \frac{1}{96}e^{\Phi}\tilde{G}^{i j k}{}_{\ell}\tilde{H}_{i j k} \pm \frac{1}{16}e^{\Phi}m L_{\ell}\big)\Gamma^{\ell}
\nonumber \\
&+& \big(\mp \frac{1}{4}\tilde{\nabla}_{\ell_1}(L_{\ell_2}) - \frac{1}{8}\tilde{\nabla}^{i}(\tilde{H}_{i \ell_1 \ell_2}) + \frac{1}{4}\tilde{\nabla}^{i}\Phi \tilde{H}_{i \ell_1 \ell_2} \pm  \frac{1}{16}h^{i}\tilde{H}_{i \ell_1 \ell_2}\big)\Gamma^{\ell_1 \ell_2}
\nonumber \\
&+& \big(\pm \frac{1}{384}e^{\Phi}\tilde{G}_{\ell_1 \ell_2 \ell_3 \ell_4}L_{\ell_5} \pm  \frac{1}{192}e^{\Phi}\tilde{H}_{\ell_1 \ell_2 \ell_3}X_{\ell_4 \ell_5}
+ \frac{1}{192}e^{\Phi}\tilde{G}^{i}{}_{\ell_1 \ell_2 \ell_3}\tilde{H}_{i \ell_4 \ell_5}\big)\Gamma^{\ell_1 \ell_2 \ell_3 \ell_4 \ell_5}\bigg)
\bigg] \eta_\pm
\nonumber \\
&+& {1 \over 2} \big(1 \mp 1\big) \bigg(h^i {\tilde{\nabla}}_i \Phi
-{1 \over 2} {\tilde{\nabla}}^i h_i \bigg) \eta_\pm \ .
\eea

Note that with the exception of the final line of the RHS of ({\ref{quad}}), all terms on the RHS of the above expression
give no contribution to the second line of (\ref{laplacian}),
using (\ref{hermiden1}), since all these terms in (\ref{quad}) are anti-Hermitian and thus the bilinears vanish.
Furthermore, the contribution to the Laplacian of $\parallel \eta_+ \parallel^2$ from the final line of ({\ref{quad}}) also vanishes;
however the final line of ({\ref{quad}}) {\it does} give a contribution
to the second line of ({\ref{laplacian}}) in the case of the
Laplacian of $\parallel \eta_- \parallel^2$.
We  proceed
to consider the Laplacians
of $\parallel \eta_\pm \parallel^2$ separately, as the analysis
of the conditions imposed by the global properties of ${\cal{S}}$
differs slightly in the two cases.

For the Laplacian
of $\parallel \eta_+ \parallel^2$, we obtain from ({\ref{laplacian}}):
\bea
\label{l1}
{\tilde{\nabla}}^{i}{\tilde{\nabla}}_{i}\parallel\eta_+\parallel^2 - (2\tilde{\nabla}^i \Phi +  h^i) {\tilde{\nabla}}_{i}\parallel\eta_+\parallel^2 = 2\parallel{\hat\nabla^{(+)}}\eta_{+}\parallel^2 - (4\kappa + 16 \kappa^2)\parallel\mathcal{A}^{(+)}\eta_+\parallel^2 \ .
\nonumber \\
\eea
This proves (\ref{maxprin}).

The Laplacian of $\parallel \eta_- \parallel^2$
is calculated from ({\ref{laplacian}}), on taking account of the contribution to the second line of
({\ref{laplacian}}) from the final line of ({\ref{quad}}). One
obtains
\bea
\label{l2}
{\tilde{\nabla}}^{i} \big( e^{-2 \Phi} V_i \big)
= -2 e^{-2 \Phi} \parallel{\hat\nabla^{(-)}}\eta_{-}\parallel^2 +   e^{-2 \Phi} (4 \kappa +16 \kappa^2) \parallel\mathcal{A}^{(-)}\eta_-\parallel^2~,
\nonumber \\
\eea
where
\bea
V=-d \parallel \eta_- \parallel^2 - \parallel \eta_- \parallel^2 h \ .
\eea
This proves (\ref{l2b}) and completes the proof.

\appendix{The geometry of ${\cal S}$}

It is known that the vector fields associated with the 1-form Killing spinor bilinears given in (\ref{1formbi}) leave invariant all the fields of
massive IIA supergravity.  In particular for massive IIA horizons we have that ${\cal L}_{K_a} g=0$ and ${\cal L}_{K_a} F=0$, $a=1,2,3$, where $F$ denotes collectively all the
fluxes of massive IIA supergravity, where $K_a$ are given in (\ref{kkk}).  Solving these conditions by expanding in $u,r$, one finds that
\begin{eqnarray}
\tilde\nabla_{(i} \tilde V_{j)}=0~,~~~\tilde {\cal L}_{\tilde V} h=\tilde {\cal L}_{\tilde V}\Delta=0~,~~~ \tilde {\cal L}_{\tilde V} \Phi=0~,
\nonumber \\
\tilde {\cal L}_{\tilde V} X=\tilde {\cal L}_{\tilde V} \tilde G=\tilde {\cal L}_{\tilde V} L=\tilde {\cal L}_{\tilde V} \tilde H=
\tilde {\cal L}_{\tilde V} S=\tilde {\cal L}_{\tilde V} \tilde F=0~.
\end{eqnarray}
Therefore $V$ is an isometry of ${\cal S}$ and leaves all the fluxes on ${\cal S}$ invariant.
Furthermore, one can establish the identities
\begin{eqnarray}
&&-2 \parallel\eta_+\parallel^2-h_i \tilde V^i+2 \langle\Gamma_+\eta_-, \Theta_+\eta_+\rangle=0~,~~~i_{\tilde V} (dh)+2 d \langle\Gamma_+\eta_-, \Theta_+\eta_+\rangle=0~,
\cr
&& 2 \langle\Gamma_+\eta_-, \Theta_+\eta_+\rangle-\Delta \parallel\eta_-\parallel^2=0~,~~~
{\tilde V}+ \parallel\eta_-\parallel^2 h+d \parallel\eta_-\parallel^2=0~,
\label{conconx}
\end{eqnarray}
which imply that ${\cal L}_{\tilde V}\parallel\eta_-\parallel^2=0$. These conditions are similar to those established for M-theory and IIA theory horizons
in \cite{11index} and \cite{iiaindex}, respectively,  but of course the dependence of the various tensors on the fields is different. In the special case that $\tilde V=0$, the
horizons are warped products of $AdS_2$ with ${\cal S}$.


\begin{thebibliography}{99}

\bibitem{carter} B. Carter, ``Black Holes,''  edited by C. de Witt and B.S. de Witt, (Gordon and Breach,
New York, 1973).



\bibitem{gibbons1} G.W. Gibbons, {\it in Supersymmetry, Supergravity and Related Topics}, eds. F. del Aguila,
J. A. de Azc`arraga and L.E. Ibanez, (World Scientific 1985).

\bibitem{gwgpkt}
  G.~W.~Gibbons and P.~K.~Townsend,
  ``Vacuum interpolation in supergravity via super p-branes,''
  Phys.\ Rev.\ Lett.\  {\bf 71} (1993) 3754
  [hep-th/9307049].


  \bibitem{maldacena}
  O.~Aharony, S.~S.~Gubser, J.~M.~Maldacena, H.~Ooguri and Y.~Oz,
  ``Large N field theories, string theory and gravity,''
  Phys.\ Rept.\  {\bf 323} (2000) 183
  [hep-th/9905111].

\bibitem{iibindex}
  U.~Gran, J.~Gutowski and G.~Papadopoulos,
  ``Index theory and dynamical symmetry enhancement near IIB horizons,''
  JHEP {\bf 1311} (2013) 104
  [arXiv:1306.5765 [hep-th]].

  \bibitem{5index}
  J.~Grover, J.~B.~Gutowski, G.~Papadopoulos and W.~A.~Sabra,
  ``Index Theory and Supersymmetry of 5D Horizons,''
  JHEP {\bf 1406} (2014) 020
  [arXiv:1303.0853 [hep-th]].

  \bibitem{11index}
  J.~Gutowski and G.~Papadopoulos,
  ``Index theory and dynamical symmetry enhancement of M-horizons,''
  JHEP {\bf 1305} (2013) 088
  [arXiv:1303.0869 [hep-th]].


  \bibitem{iiaindex}
  U.~Gran, J.~Gutowski, U.~Kayani and G.~Papadopoulos,
  ``Dynamical symmetry enhancement near IIA horizons,''
  arXiv:1409.6303 [hep-th].






  \bibitem{isen}
J. Isenberg and V. Moncrief, \textit{Symmetries of cosmological Cauchy horizons,
Commun. Math. Phys.} {\bf{89}} (1983) 387.

\bibitem{gnull}
H. Friedrich, I. Racz and R. M. Wald,
\textit{On the rigidity theorem for space-times with a stationary event horizon or a compact Cauchy horizon,
Commun. Math. Phys.} {\bf{204}} (1999) 691; [gr-qc/9811021].





  \bibitem{romans}
  L.~J.~Romans,
  ``Massive N=2a Supergravity in Ten-Dimensions,''
  Phys.\ Lett.\ B {\bf 169} (1986) 374.


\bibitem{roo}
  E.~A.~Bergshoeff, M.~de Roo, S.~F.~Kerstan, T.~Ortin and F.~Riccioni,
  ``IIA ten-forms and the gauge algebras of maximal supergravity theories,''
  JHEP {\bf 0607} (2006) 018
  [hep-th/0602280].






   \bibitem{index}
M. F. Atiyah and I. M. Singer,
  ``The Index of elliptic operators. 1"
  Annals Math.\  {\bf 87} (1968) 484.













\end{thebibliography}
\end{document}